\documentclass[aps, pre, twocolumn, nofootinbib]{revtex4-2}
\usepackage[utf8]{inputenc}
\usepackage{csquotes}
\pdfoutput=1

\usepackage{graphicx}% Include figure files
\graphicspath{ {figures/} }
\usepackage{dcolumn}% Align table columns on decimal point
\usepackage{bm}% bold math
\usepackage{amsmath}
\usepackage{amssymb}
\usepackage{mathrsfs}
\usepackage{siunitx}
\usepackage{xurl}

\usepackage{tabularx}
\usepackage{microtype}
\usepackage{hyperref}
\usepackage{multirow}
\usepackage{float}

\makeatletter
\newcommand*{\balancecolsandclearpage}{%
  \close@column@grid
  \clearpage
  \twocolumngrid
}
\makeatother

\hypersetup{
    colorlinks=true,
    linkcolor=blue,
    filecolor=magenta,      
    urlcolor=cyan,
}

\makeatletter
\def\l@subsubsection#1#2{}
\renewcommand{\p@paragraph}{}
\makeatother
\NewDocumentCommand{\dn}{e{_^}}{%
  _{\IfValueT{#1}{#1}\vphantom{\smash[b]{|}}}
  ^{\IfValueT{#2}{#2}\vphantom{\smash[t]{\big|}}}
}

\begin{document}

\title{%Alternative titles:\\
The foreshadow of a second wave:\\
An analysis of current COVID-19 fatalities in Germany
%No apparent discrepancy between observed COVID-19 infections and fatalities
}

\author{
Matthias Linden$^{1,2}$,
Jonas Dehning$^{1}$,
Sebastian B. Mohr$^{1}$,
Jan Mohring$^{3}$,
Michael Meyer-Hermann$^{4,5}$,
Iris Pigeot$^{6,7}$,
Anita Schöbel$^{5,8}$,
Viola Priesemann$^{1,9,*}$% $^{1,2,3,4,5,6,7,8}$
}
\affiliation{\vspace{0.15cm}
$^1$ \mbox{Max Planck Institute for Dynamics and Self-Organization, Am Fassberg 17, 37077 G\"ottingen.}\\
$^2$ \mbox{Institute for Theoretical Physics, Leibniz University, 30167 Hannover.}\\
$^3$ \mbox{Fraunhofer Institute for Industrial Mathematics, Fraunhofer-Platz 1, 67663 Kaiserslautern.}\\
$^4$ \mbox{Department of Systems Immunology and Integrated Centre of Systems Biology (BRICS),} \mbox{Helmholtz Centre for Infection Research,  Rebenring 56, 38106 Braunschweig.}\\
$^5$ \mbox{Institute for Biochemistry, Biotechnology and Bioinformatics, Technische Universität Braunschweig, 38106 Braunschweig.}
$^6$ \mbox{Leibniz Institute for Prevention Research and Epidemiology – BIPS, Achterstraße 30, 28359 Bremen.}\\
$^7$ \mbox{University Bremen, Faculty 03: Mathematics/Computer Science,} Bibliothekstraße 5, 28359 Bremen.\\
$^8$ \mbox{Fachbereich Mathematik, University Kaiserslautern, Postfach 3049, 67663 Kaiserslautern.}\\
$^9$ \mbox{Department of Physics, University of Göttingen, Friedrich-Hund-Platz 1, 37077 Göttingen.}\vspace{0.15cm}
$^{*}$\mbox{Correspondence should be addressed to \href{mailto:viola.priesemann@ds.mpg.de}{viola.priesemann@ds.mpg.de}}\\
%A German version of the manuscript is available on \url{insert.link}!
A German version of the manuscript is appended after supplemental material.
}

\date{\today}

\maketitle

% Currently about 750 words (excluding figure caption; 750 are allowed (I dont know whether with or without figure caption.

% Max 5 references!!

% Deutsches Ärzteblatt, Kurzmitteilung  (research  letter):  Diese  Rubrik  eignet  sich  für  Originaldaten,  die  in  komprimierter  Form  darstellbar  sind  (Textumfang:  750 Wörter, 2 kompakte Tabellen oder Grafiken, bis zu 5 Referenzen, keine Zusammenfassung):   https://www.aerzteblatt.de/down.asp?typ=PDF&id=2414

%\subsection*{Introduction}

A second wave of SARS-CoV-2 is unfolding in dozens of countries. Compared to the first wave, this second wave manifests itself strongly in newly reported cases, but less so in death counts. However, already the increasing case numbers put the mitigation strategy at risk, because with case numbers being too high, local tipping points are crossed, which makes the control increasingly difficult, impedes the targeted protection of the vulnerable people, and leads to self-accelerating spread~\cite{contreras_challenges_2020,Meyer-Hermann_Together_2020}.

Importantly, a tipping point is reached when case numbers surpass the test-trace-and-isolate (TTI) capacity of the local health authority, thus potentially long before hospitals are overwhelmed~\cite{contreras_challenges_2020}: 
%When TTI breaks down, tests results are obtained only with considerable delay, and TTI misses more and more chains of infections. 
When TTI breaks down, testing and tracing can only be carried out insufficiently and with considerable delay, which leads to more and more missed chains of infections. Thereby increasingly more SARS-CoV-2 carriers remain unnoticed, and thus transmit the virus inadvertently -- also to people at risk. As unaware carriers are strong drivers of the spread of SARS-CoV-2, the case numbers start to rise more quickly, and the spread spills over to the vulnerable population. In the following, we present evidence for a second wave and the crossing of this tipping point. %~\cite{noauthor_gesundheitsamter_nodate,noauthor_zwei_nodate}.

%\subsection*{Methods}

\begin{figure}[H]
    \includegraphics[width=0.48\textwidth]{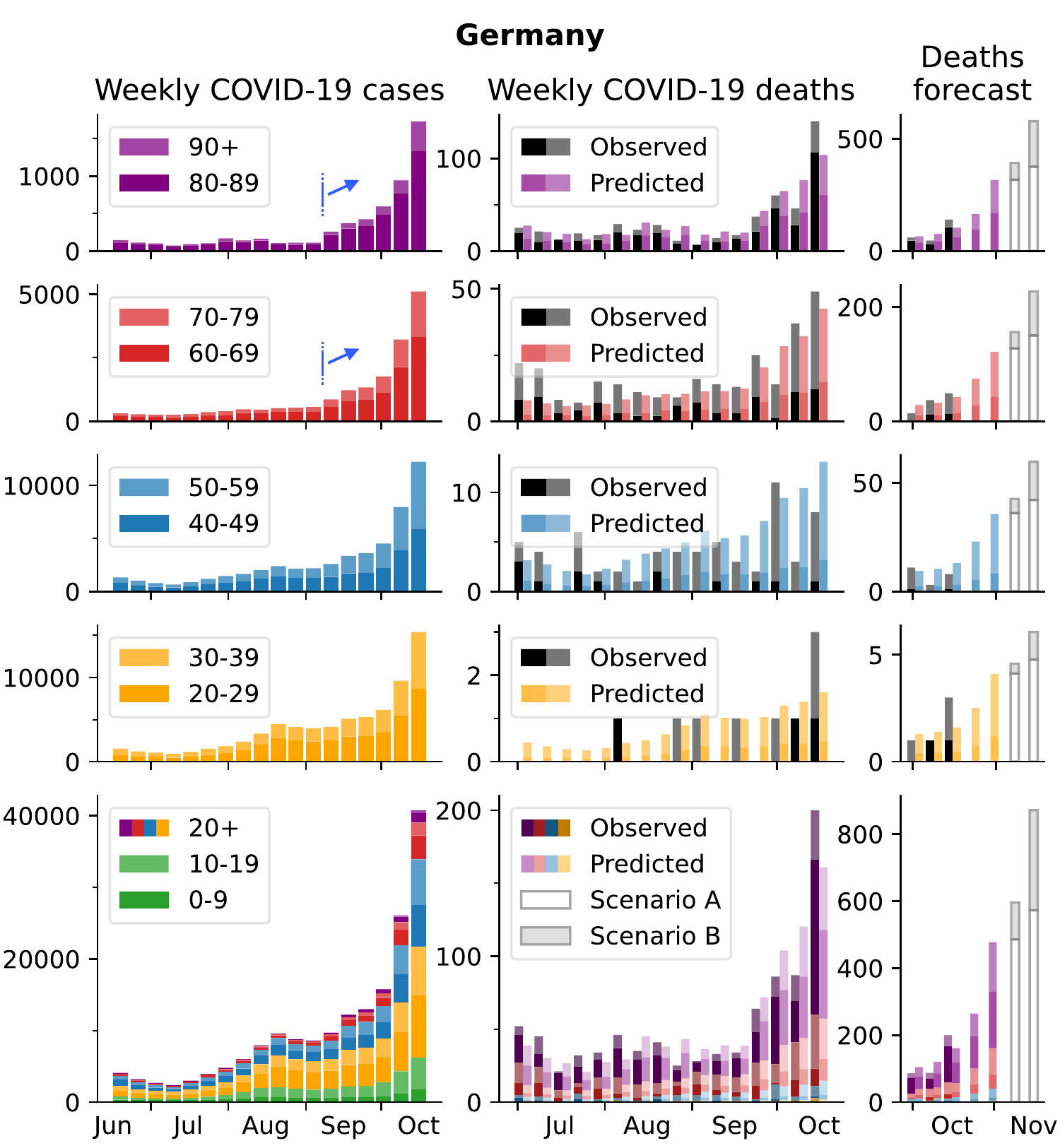}
    \caption{The number of weekly COVID-19 cases (left), and the number of predicted and actual COVID-19-associated deaths are displayed for each age group (middle). The observed number deaths (black) in each age group matches well the predicted deaths calculated from the case numbers (color) using an age-dependent infection-fatality rate from a meta-analysis\cite{Levin2020}. 
    Despite a rise in total reported cases since mid July (bottom), cases in the older age groups only started to rise in September (blue arrows). As a consequence, a rise in deaths is expected to unfold soon (right). The forecast scenarios A and B (gray, right column) predict a further increase in case numbers and deaths. The recent rise of infections in the older age groups together with the overall rise in cases clearly indicates a loss of control over the spreading and the start of a second wave.
    % From the age distribution of cases an effective infection-fatality ratio $\mathrm{IFR}_\mathrm{eff}$ can be calculated, which is currently at 0.75\% in Germany.
    }
    \label{fig:DE}
\end{figure}

In Germany, as in many other countries, the emergence of the second wave has been questioned, because only the case numbers have been rising since July, while hospitals were not overwhelmed and death numbers stayed almost constant
(Fig.~\ref{fig:DE}).
 We investigated this apparent discrepancy using age-stratified case and death reports \cite{RKI_situation_reports}, and an age-dependent infection fatality rate (IFR). The IFR is derived from a meta-analysis of seroprevalence studies and extensive PCR tracing programs \cite{Levin2020}. Approximately, the IFR increases by a factor of 10 for every 20 years of age, being about 10\% at age 82. From this age-dependent IFR we predict the temporal evolution of the COVID-19-associated deaths by delaying each age group’s observed weekly cases by two weeks and multiplying by the IFR (see supplementary material). 

%\subsection*{Results}

The temporal evolution of the predicted COVID-19-associated deaths (color) matches the actually observed ones (black) well in each age group (Fig.~\ref{fig:DE}, right). Even the absolute numbers match, suggesting a low fraction of unobserved SARS-CoV-2 infection chains.
This age-stratified analysis plausibly resolves the apparent discrepancy between the total reported infections and deaths: The rise in reported cases has been mainly driven by younger generations (younger than 60), who contribute only little to the absolute death counts, whereas the elderly (60+ years) managed to maintain low case numbers until recently. However, in September cases among the elderly started to increase steeply. This finding is not confounded by increased testing, because the test numbers in that age group stayed almost constant~\cite{RKI_situation_reports}. % Fig. 5 and 6 in RKI_LabSurveillance
This spill-over to the elderly strongly suggests a failure to protect the vulnerable people due to an increased number of unaware SARS-CoV-2 carriers. Moreover, this loss of control over the spread is also marked by the increase of the reproduction number R since begin of October (Table~\ref{tab:IFR_CFR}). The spill-over and the increase of R are clear signatures of crossing the tipping point to uncontrolled spread.

During summer 2020, the age groups 60+ apparently were protected well:
If the virus had been spreading independently equally among all age groups, an estimated 1.3\% to 2.0\% of the infected would have died (calculated from the IFR in \cite{Levin2020}). However, the \textit{case fatality rate} CFR was only  0.39\% in August (Table~\ref{tab:IFR_CFR}). 
Now, with the strongly increasing number of infections among the elderly (Fig.~\ref{fig:DE}), the CFR will most likely increase soon. This indicates that protecting the people at risk might be possible whilst case numbers are low, but likely fails under high COVID-19 incidence.

%\subsection*{Discussion and Outlook}

For the coming two weeks, the already reported cases clearly predict that the number of weekly deaths will almost double. Beyond two weeks forecasts are difficult, because the death counts depend on the not-yet-reported cases. To account for this uncertainty, we display two future scenarios, where in (A) we assume a linear increase, and in (B) an exponential increase in cases (see supplementary material). The resulting forecasts range between 500 and 800 weekly deaths in early November, but case numbers might even grow faster in case the second wave fully unfolds.

If the test-trace-and-isolate (TTI) system is overwhelmed and stops to function as a powerful means of control, mitigation of the spread breaks down and the rise of case numbers accelerates~\cite{contreras_challenges_2020}.
Thus, the current development in Germany puts the past successful management of the pandemic seriously at risk. To re-establish control and to avoid crossing the tipping point when the TTI capacity is exceeded, case numbers have to be lowered without further delay. Otherwise the control of the spread and the protection of vulnerable people will inevitably require much more restrictive measures -- at latest when the hospital capacity is reached.

\begin{table}[H]
    \centering
    \vspace{0.7cm}
    \begin{tabular}{l|l|l|l|l||l|l}
    \multicolumn{2}{r|}{Number of}&\multicolumn{3}{c||}{Number of} &\multicolumn{2}{c}{Metrics}\\
    \multicolumn{2}{r|}{Weeks}&Cases&\multicolumn{2}{c||}{Deaths}&CFR &$ R$\\
Period&&&pred.&obs.&[\%]&(RKI) \\ \hline
Jul&4&14163 & 122.5 & 146 &1.0&1.11\\
Aug&4&32273 & 156.3 & 127 &.39&1.06\\
early Sep&3& 30493 & 214.4 & 184 &.6&1.10\\
late Sep&2& 28737 & 281 & 286 & .99&1.16\\
early Oct&2& 66796 & 741 &- & - & 1.31\\
    \end{tabular}
    \caption{\textbf{This table summarizes the data presented in Fig.~\ref{fig:DE}.}  For each time-period in summer 2020, the number of COVID-19 infections (“cases”) are displayed together with the number of deaths as observed (obs.) and as predicted (pred.) from the age-dependent IFR \cite{Levin2020}. For correct association between case and death, the number of deaths have been advanced by 2 weeks. Predicted and observed deaths deviate less than 20\% from each other. As metrics, the case fatality rate (CFR) is calculated as ratio between observed deaths and cases. For comparison: If infections were evenly distributed across the population and no infection remained undetected, the CFR would be equal to the theoretical infection fatality rate (IFR) for the German population, i.e. 1.61 with a 95\% confidence interval of [1.29;2.02]~\cite{Levin2020}. The last column displays the reproduction number R as reported by the Robert Koch Institute (RKI).}

    %Comparison of predicted and observed deaths.} This table sums up data presented in Fig.~\ref{fig:DE}. Predicted and observed deaths are associated with the assumed date of reporting of the infection 2 weeks earlier. The effective IFR is calculated from the age-distribution of cases using an age-dependent IFR\cite{Levin2020}.  The CFR is calculated from the observed deaths. Predicted and observed deaths are within 20\% of each other. Both the IFR and the reproductive number $R$ (as published by RKI) increase since early September.}% \textit{early Oct} contains data from the first two weeks. \textit{late Sep} starts 2020/09/21 containing two weeks and \textit{early Sep} contains the first three weeks of September. }%For comparison: $\mathrm{IFR}_\textrm{Levin}^\mathrm{DE}=1.61 [1.29; 2.02]$ would be observed as CFR if infections were evenly distributed across the population and no infections remained undetected.}
    \label{tab:IFR_CFR}
\end{table}

\balancecolsandclearpage
\section*{Supplementary material}
%\section{supplementary methods section}
\paragraph{Data sources}

For Germany, the age-stratified data used in this document was retrieved from the national health agency, the Robert Koch Institute (RKI): both the number of reported SARS-CoV-2 cases and the number of deaths are available in age groups of ten years  \cite{RKI_situation_reports}, and the reported cases are also available with 5-year stratification up to the age of 80 from \cite{SurvStat}. It was verified that the number of cases in the two sources matches. The time points of the case and death numbers correspond to their \textit{reporting} dates, and not necessarily to the actual testing or date of death.

The population age distribution of Germany was retrieved from the United Nation's database \cite{data-Pop}. 
 
\paragraph{IFR calculation}

The overall goal is to estimate death numbers from past reported cases per age group and compare them to the observed number of deaths. 
We used an age-dependent infection-fatality rate (IFR) equation, following recent meta-studies \cite{Levin2020,Moraga2020}: 
Let $a$ be the age of an infected person. Then the infection fatality rate for this age is estimated as
\begin{align}
\textrm{IFR}(a) = 0.1 \cdot 10\dn_{}^{\frac{1}{20}(a-82)}
\label{eqn:IFR}
\end{align}
This equation formalizes that at 82 years, the IFR is 10\% and decreases by a factor of 10 every 20 years. The age group 100+ is treated separately as 60\%, in accordance with the supplementary calculations in \cite{Levin2020}. 
In our analysis, we omit cases below the age of 20, as the weekly reported death numbers in that group are too low for any meaningful analysis.
The equation \eqref{eqn:IFR} is very close to the published one \cite{Levin2020}, but has been rounded to integers for simplicity. For comparison see table \ref{tab:IFR_numbers}.

As reported cases are available with 5-year stratification, equation \eqref{eqn:IFR} was evaluated centered in the respective age group. For instance, the IFR of the age group 40-44 is approximated as:
\begin{align}
\textrm{IFR}^{40-44} = \textrm{IFR}(42)  = 0.1 \cdot 10\dn_{}^{\frac{1}{20}(42-82)} = 0.1\%
\label{eqn:IFR_example}
\end{align}

One has to distinguish this IFR calculation from the CFR calculation which we calculate by simply dividing the number of deaths by the observed cases two weeks before, as discussed in more detail in paragraph \ref{para_CFR}.

\paragraph{Estimating the number of deaths from the reported SARS-CoV-2 cases}\label{sec:est_deaths}

The number of deaths is estimated by multiplying the published weekly number of reported cases in 5-years-wide age groups by the associated IFR (equation \eqref{eqn:IFR_example}). To obtain the time point of the expected number of deaths, we delayed the number of reported cases by two weeks. This accounts for a one week delay between infection and case reporting, and a three weeks delay between infection and death. Thus, the total delay of two weeks between reported case and death amounts to the difference between the infection-case and infection-death delay. In addition, two weeks matches a median delay of 10 days, which can be derived from individual case histories supplied by the RKI's publicly available database \cite{arcgis}.

An alternative approach would have been to take those case histories, infer a delay-distribution between reporting and death with a one-day temporal resolution and predict daily deaths. Unfortunately, the age-groups of this dataset are between 20 and 25 years wide, compared to 5 year groups with weekly temporal resolution \cite{SurvStat}. Due to the strong age-specificity of the IFR we opted for smaller age-groups.

The observed number of deaths is only available in 10-years-wide age groups. Thus, to match the coarser stratification, we summed the corresponding two 5-years-wide age groups for all the plots.

\paragraph{Uncertainties of predictions}

Our estimations features a number of uncertainties: The IFR probably decreased over the course of the pandemic because of better treatments. This leads to some overestimation of the IFR as reported by Levin et al \cite{Levin2020}, which uses data from early in the pandemic. A fraction of cases remains unobserved, and people at risk may protect themselves better than earlier in the pandemic. Moreover, the assumed two-week delay between case detection and death is an approximation, the observed delay of individual cases features a wider distribution. Overall however, the evidence indicates that in Germany half or more of all cases are currently detected \cite{Russell2020.07.07.20148460}. In relative numbers across age-groups, uncertainties are smaller, because systematic errors in the IFR and the delay approximations are expected to effect all age groups similarly.

\paragraph{Accounting for non-matching age groups}

If reported cases are not available with a 5-year stratification, but only coarser , we distribute the reported cases of the coarser age groups in 5-years-wide age groups, proportional to the population-age distribution of the respective country \cite{data-Pop}. 
This approach is necessary for ages above 80 years for Germany, where only 10-years-wide age groups are available: 80-89, 90-99 and 100+.

\paragraph{Prediction}
We consider two distinct scenarios for the future spread of the virus. 
\emph{Scenario A} is a linear extrapolation fitted to the reported cases in the respective age groups over the last 5 weeks, which we chose since linear growth can reflect the local and heterogeneous spreading of SARS-CoV-2.
Furthermore, \emph{Scenario B} is an exponential forecast using the reproduction rate $R_0=1.2$ given by the RKI \cite{RKI_situation_reports} in early October and represents a worser scenario. For the forecast, the corresponding weekly increase of 37.5\% is applied to the latest week of age-stratified case numbers.

\paragraph{Effective infection fatality rate}
The effective infection fatality rate $\textrm{IFR}_\textrm{eff}$ is defined as the infection fatality rate given the age distribution of reported cases. It is calculated by weighting the IFR of the age groups (eq.~\eqref{eqn:IFR_example}) by the number of current reported cases.
This approach to calculate the effective IFR accounts for varying case distributions within the age groups, which therefore leads to a time-dependence of the effective IFR. Note that this $\textrm{IFR}_\textrm{eff}$ is not a case-fatality rate, because only the age distribution of reported cases is used to weight the IFR$(a)$ in each age group appropriately; this is different from the case-fatality rate, which corresponds to the ratio of observed deaths to reported cases. 

We used this definition of an effective IFR to calculate \emph{the} effective IFR for Germany assuming eq.~\eqref{eqn:IFR_example} and a case distribution proportional to the population's age distribution: $\mathrm{IFR}_\textrm{mix}^\mathrm{DE}=1.83$. With equal spread across all age groups, 65\% of deaths would originate from the 80+ age group (Tabl.~\ref{tab:IFR_numbers}).
With values from \cite{Levin2020} for the age-specific IFR (instead of eq.
~\eqref{eqn:IFR_example}), the corresponding effective IFR for Germany is $\mathrm{IFR}_\textrm{Levin}^\mathrm{DE}=1.61 [1.29; 2.02]$. Both $\mathrm{IFR}^\mathrm{DE}$ represent an age-agnostic equal infection spread.

\paragraph{Case fatality rate}
\label{para_CFR}

In order to compare the theoretical effective infection fatality rate to the actual reported deaths, we calculated a case fatality rate. The approach was to divide the number reported deaths occurring two weeks later by the reported cases. This doesn't take into account the wide distribution of  delays between infections and deaths. We mitigate the errors introduced by this simple assumption by using at least 2 weeks of data (Tab.~\ref{tab:IFR_CFR}).

\paragraph{Testing}
The steep increase in the number of cases throughout all age-groups starting in October is not driven by increased testing activity: Between the end of June and the end of September, the rate of positive tests stayed below 1\% except for the age group 15-34, which stayed between 1-1.5\% \cite{RKI_LabSurveillance}. Until mid-September, it even remained below 0.5\% in the age groups 60-79 and 80+. In conjunction with a consistently high number of tests of 600-700 per 100000 in the age group 80+, well above the 300-400 tests per 100000 in the other groups, we assume adequate coverage of that age group, and as a result a constant low number of undetected cases. The twofold increase in overall testing from beginning to end of August, 500000 to 1 million tests per week \cite{RKI_situation_reports}, was driven by tests in the age groups $<60$, temporarily lowering the fraction of positive tests in those groups until mid-September. From then on the number of tests continues to be around 1 million tests per week, leading to our conclusion that the increase of cases since September is not driven by increased test activity.

\begin{table}[H]
    \centering
    \vspace{0.7cm}
    \begin{tabular}{cc|ll|l|l|l|l}
\multicolumn{6}{c|}{}&\multicolumn{2}{c}{cum. [\%]}\\
\multicolumn{2}{l|}{}&\multicolumn{3}{c|}{IFR (in percent)}&[\%]&\multicolumn{2}{c}{of deaths}\\
\multicolumn{2}{l|}{Age group}&\multicolumn{2}{c|}{Levin et al.}&Eq.~\eqref{eqn:IFR_example}&of pop.&pred.&obs.\\\hline
\parbox[t]{2mm}{\multirow{4}{*}{\rotatebox[origin=c]{90}{20-39}}}&20-24&.007&.006 .009]&.01&5.43&.0\\
&25-29&.014&[.011 .017]&.0178&5.76&.126&.24\\
&30-34&.025&[.021 .03]&.0316&6.50&.23\\
&35-39&.045&[.038 .053]&.0562&6.48&.43&1.12\\\hline
\parbox[t]{2mm}{\multirow{4}{*}{\rotatebox[origin=c]{90}{40-59}}}&40-44&.081&[.07 .094]&.1&6.04&.76\\
&45-49&.148&[.129 .17]&.178&6.19&1.36&2.58\\
&50-54&.269&[.234 .308]&.316&7.97&2.74\\
&55-59&.488&[.425 .561]&.562&8.12&5.23&10.1\\\hline
\parbox[t]{2mm}{\multirow{4}{*}{\rotatebox[origin=c]{90}{60-79}}}&60-64&.887&[.764 1.03]&1.&6.95&9.03&\\
&65-69&1.61&[1.37 1.9]&1.78&5.76&14.6&20.2\\
&70-74&2.93&[2.44 3.52]&3.16&5.58&22.5\\
&75-79&5.33&[4.34 6.53]&5.62&4.34&35.9&43.0\\\hline
\parbox[t]{2mm}{\multirow{5}{*}{\rotatebox[origin=c]{90}{80+}}}&80-84&9.68&[7.71 12.2]&10.&3.89&57.1\\
&85-89&17.6&[13.7 22.6]&17.8&1.95&76.1&83.1\\
&90-94&32.&[24.2 42.2]&31.6&.90&91.8\\
&95-99&&&56.2&.24&99.2&99.3\\
&100+&&&60.&.02&100.&100.\\ 
    \end{tabular}
    \caption{
    \textbf{Comparison of infection fatality rates (IFR).} The first column reproduces the age-dependent IFR from Levin et al.~\cite{Levin2020} evaluated at the center of the age group alongside 95\% confidence interval bounds. The second column list the IFR values used in the analysis, which are calculated using eq.~\eqref{eqn:IFR_example}. 
    The right half displays the predicted and observed percentage of deaths in each age group, cumulatively summing over all preceding age groups. For the estimated percentage for Germany, the age-specific IFR from the 3rd column is multiplied with the fraction of population in the age group. The observation refers to deaths in weeks 31-40 \cite{RKI_situation_reports}.
    The observed fraction of deaths was almost twice as large as the estimated one in the age group up to 60 years (last two columns), reflecting that the virus spread more among the younger people than expected. These columns also show that 5.23 \% (10.1\%) of all deaths are expected among the age group up to 60, thus almost 95 \% (90 \%) of all deaths are expected to originate from the age groups 60+ assuming age-independent (or age-dependent) spreading.
    }
    \label{tab:IFR_numbers}
\end{table}

%\clearpage
%Force figures/tables above references
\bibliography{references}

\begin{thebibliography}{10}

\bibitem{contreras_challenges_2020}
Sebastian Contreras, Jonas Dehning, Matthias Loidolt, F.~Paul Spitzner,
  Jorge~H. Urrea-Quintero, Sebastian~B. Mohr, Michael Wilczek, Johannes
  Zierenberg, Michael Wibral, and Viola Priesemann.
\newblock The challenges of containing {SARS}-{CoV}-2 via
  test-trace-and-isolate.
\newblock {\em arXiv:2009.05732 [q-bio]}, September 2020.
\newblock arXiv: 2009.05732.

\bibitem{Meyer-Hermann_Together_2020}
Michael Meyer-Hermann, Iris Pigeot, Viola Priesemann, and Anita Schöbel.
\newblock Together we can do it: {Each} individual contribution protects
  health, society, and the economy,
  \url{https://www.mpg.de/15503604/statement-non-university-research-organizations-covid-19-epidemic.html}.
\newblock 2020.

\bibitem{RKI_situation_reports}
{RKI} {Tuesdays'} situation reports, retrieved 2020/10/07,
  \url{https://www.rki.de/DE/Content/InfAZ/N/Neuartiges_Coronavirus/Situationsberichte/Gesamt.html}.

\bibitem{Levin2020}
Andrew~T Levin, William~P. Hanage, Nana Owusu-Boaitey, Kensington~B. Cochran,
  Seamus~P. Walsh, and Gideon Meyerowitz-Katz.
\newblock Assessing the age specificity of infection fatality rates for
  {COVID-19}: Systematic review, meta-analysis, and public policy implications.
\newblock {\em medRxiv}, 2020.
\newblock medRxiv: 10.1101/2020.07.23.20160895v5.

\bibitem{SurvStat}
Robert Koch-Institut (RKI).
\newblock {SurvStat@RKI} 2.0 {API}.

\bibitem{data-Pop}
United nations world population database,
  \url{https://population.un.org/wpp/Download/Standard/Population/}.

\bibitem{Moraga2020}
P~Moraga, DI~Ketcheson, HC~Ombao, and CM~Duarte.
\newblock Assessing the age- and gender-dependence of the severity and case
  fatality rates of {COVID-19} disease in {Spain} [version 1; peer review: 1
  approved].
\newblock {\em Wellcome Open Research}, 5(117), 2020.

\bibitem{arcgis}
Robert Koch-Institut~(RKI) dl-de/by 2-0.
\newblock {ArcGIS} database of the {RKI}'s {COVID-19} dashboard,
  \url{https://www.arcgis.com/home/item.html?id=f10774f1c63e40168479a1feb6c7ca74}.

\bibitem{Russell2020.07.07.20148460}
Timothy~W Russell, Nick Golding, Joel Hellewell, Sam Abbott, Lawrence Wright,
  Carl A~B Pearson, Kevin van Zandvoort, Christopher~I Jarvis, Hamish Gibbs,
  Yang Liu, Rosalind~M Eggo, John~W Edmunds, and Adam~J Kucharski.
\newblock Reconstructing the early global dynamics of under-ascertained
  covid-19 cases and infections.
\newblock {\em medRxiv}, 2020.

\bibitem{RKI_LabSurveillance}
{RKI Laborbasierte Surveillance SARS-CoV-2} {report of 2020-09-30} retrieved
  2020/10/06, \url{https://ars.rki.de/Content/COVID19/Main.aspx}.

\end{thebibliography}
\bibliographystyle{unsrt} 

\balancecolsandclearpage
\section*{Haupttext}
%\subsection*{Einleitung}

Eine zweite Welle von SARS-CoV-2 breitet sich in Dutzenden von Ländern aus. Die zweite Welle zeigt sich zwar stark in neu gemeldeten Fällen, aber im Vergleich zur ersten Welle weniger in der Zahl der Todesfälle. Jedoch gefährden bereits diese steigenden Fallzahlen die Eindämmung von COVID-19, denn bei zu hohen Fallzahlen werden lokal Kipppunkte überschritten, was die Kontrolle erschwert, den gezielten Schutz von Risikogruppen behindert und zu einer selbstbeschleunigenden Ausbreitung führt~\cite{contreras_challenges_2020,Meyer-Hermann_Together_2020}.

Ein Kipppunkt wird erreicht, wenn die täglichen Neuinfektionen die Kapazität einer lokalen Gesundheitsbehörde übersteigen, so dass sie nicht mehr effektiv testen und Kontakte nachverfolgen können (TTI, test-trace-isolate). Diese Situation kann eintreten lange bevor die Krankenhäuser überlastet sind, und ist daher besonders kritisch~\cite{contreras_challenges_2020}: 
Wenn das TTI zusammenbricht, können Tests und Rückverfolgung nur unzureichend oder mit großen Verzögerungen durchgeführt werden, wodurch Infektionsketten zunehmend nicht erkannt werden. Dadurch bleiben mehr SARS-CoV-2-Träger unentdeckt und können das Virus unbeabsichtigterweise übertragen -- auch auf Risikogruppen. Da die unentdeckten Träger die Ausbreitung entscheidend vorantreiben, steigen die Fallzahlen immer schneller und Ansteckungen breiten sich auch in vorher effektiv geschützten Risikogruppen aus. Im Folgenden präsentieren wir Hinweise für eine zweite Welle und die Überschreitung dieses Kipppunktes.

%\subsection*{Methoden}

\begin{figure}[ht!]
    \includegraphics[width=0.48\textwidth]{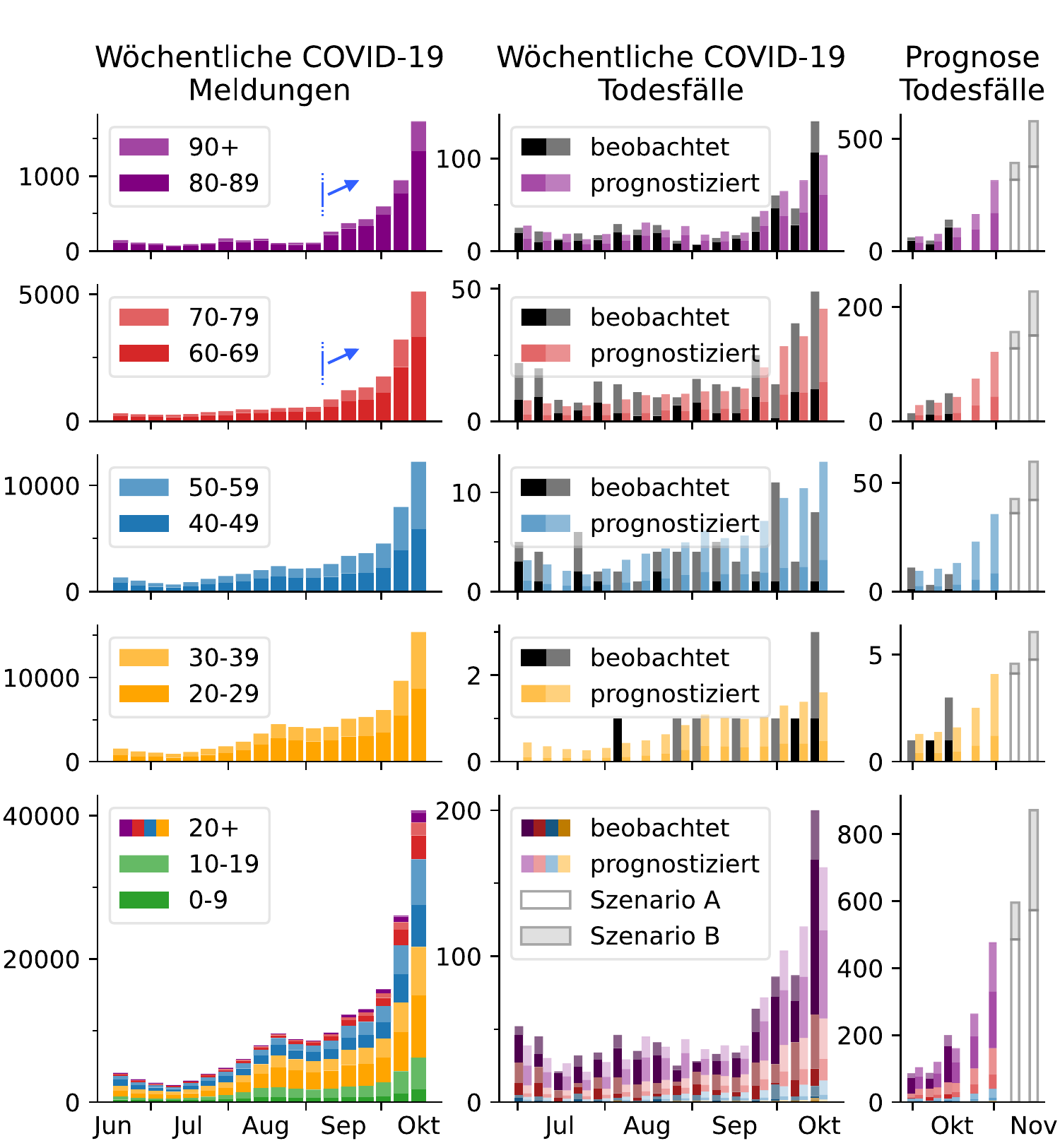}
    \caption{Die Anzahl der wöchentlichen COVID-19-Fälle (links) sowie die Anzahl der prognostizierten und beobachteten COVID-19-assoziierten Todesfälle werden für jede Altersgruppe (Mitte) angezeigt. Die Vorhersage der Anzahl der Todesfälle anhand der Fallzahlen (farbig) stimmt mit den gemeldeten Todesfällen (schwarz) in jeder Altersgruppe gut überein, wobei eine altersabhängige Infektionsfatalitätsrate basierend auf einer Meta-Analyse~\cite{Levin2020}
    zugrunde gelegt wurde. Trotz eines Anstiegs der Gesamtzahl der gemeldeten Fälle seit Mitte Juli (unten) stiegen die Fälle in den älteren Altersgruppen erst seit September an (blaue Pfeile). Infolgedessen erwarten wir, dass bald ein deutlicher Anstieg der Todesfälle beobachtet werden kann. Die Prognoseszenarien A und B (grau, rechte Spalte) sagen einen weiteren Anstieg der Fallzahlen und Todesfälle voraus. Der jüngste Anstieg der Infektionen in den älteren Altersgruppen zusammen mit dem Gesamtanstieg der Fälle weist eindeutig auf einen Kontrollverlust über die Ausbreitung und den Beginn einer zweiten Welle hin
    }
    \label{fig:DE_DE}
\end{figure}

In Deutschland, wie in vielen anderen Ländern, ist die Entstehung einer zweiten Welle in Frage gestellt worden, da nur die Fallzahlen seit Juli gestiegen sind, während die Krankenhäuser nicht überlastet waren und die Todesfälle nahezu konstant blieben
(Abb.~\ref{fig:DE_DE}).
Wir untersuchten diese offensichtliche Diskrepanz anhand von altersstratifiziertn Fall- und Sterbezahlen \cite{RKI_situation_reports} und einer altersabhängigen Infektionssterblichkeitsrate (IFR, infection fatality rate). Die IFR basiert auf einer Metaanalyse von Seroprävalenzstudien und umfangreichen PCR-Tracing-Programmen \cite{Levin2020}.
Die IFR steigt für jedes 20. Lebensjahr etwa um den Faktor  und beträgt im Alter von 82 Jahren rund 10\%. Aus dieser altersabhängigen IFR prognostizieren wir die zeitliche Entwicklung der COVID-19-assoziierten Todesfälle, indem wir die wöchentlich beobachteten Fälle jeder Altersgruppe um zwei Wochen verzögern und mit der entsprechenden IFR multiplizieren. Weitere Informationen und Ergebnisse finden sich in den ``supplementary material''.

%\subsection*{Ergebnisse}

Der zeitliche Verlauf der vorhergesagten COVID-19-assoziierten Todesfälle (farbig) stimmt mit den tatsächlich beobachteten (schwarz) in jeder Altersgruppe gut überein (Abb.~\ref{fig:DE_DE}, rechts). Selbst die absoluten Zahlen stimmen überein, was auf einen niedrigen Anteil unbeobachteter SARS-CoV-2-Infektionsketten hindeutet. 

Unsere altersstratifizierte Analyse löst die scheinbare Diskrepanz zwischen den insgesamt gemeldeten Infektionen und Todesfällen plausibel auf: Der Anstieg der gemeldeten Fälle wurde hauptsächlich von den jüngeren Generationen (jünger als 60 Jahre) verursacht, die nur wenig zu den absoluten Todesfällen beitragen, während es bei den älteren Menschen (60+ Jahre) bis vor Kurzem gelang, die Fallzahlen niedrig zu halten. Im September stiegen die Fälle bei älteren Menschen jedoch steil an. Diese Beobachtung lässt sich nicht auf die Zunahme der durchgeführten Tests zurückführen, da die Anzahl der Tests in dieser Altersgruppe nahezu konstant blieb~\cite{RKI_situation_reports}.
Dieses Überschwappen auf ältere Menschen legt nahe, dass es aufgrund der wachsenden Zahl unbemerkter SARS-CoV-2 Infizierter nicht gelungen ist, die Risikogruppen zu schützen. Darüber hinaus ist dieser Kontrollverlust seit Anfang Oktober auch durch einen Anstieg der Reproduktionszahl $R$ gekennzeichnet (Tabelle~\ref{tab:IFR_CFR_DE}). Das Überschwappen auf die ältere Generation und die Zunahme in $R$ sind beides klare Zeichen dafür, dass der Kipppunkt zur unkontrollierten Ausbreitung überschritten wurde.

Im Sommer 2020 waren die Altersgruppen 60+ offenbar gut geschützt: Hätte sich das Virus unabhängig und gleichmäßig über alle Altersgruppen verbreitet, wären schätzungsweise 1,3\% bis 2,0\% der Infizierten gestorben, wenn man von der in \cite{Levin2020} berichteten IFR ausgeht. Die \textit{Todesfallrate} (CFR, case fatality rate) lag im August jedoch nur bei 0,39\% (Tabelle~\ref{tab:IFR_CFR_DE}). 
Jetzt, mit der stark zunehmenden Anzahl der Infektionen unter älteren Menschen (Abb.~\ref{fig:DE_DE}), wird die CFR höchstwahrscheinlich bald ansteigen. Dies weist darauf hin, dass ein Schutz der Risikopersonen möglich ist, solange die Fallzahlen niedrig sind, aber wahrscheinlich bei hoher COVID-19-Inzidenz scheitert.

%\subsection*{Diskussion und Ausblick}

Für die kommenden zwei Wochen kann alleine aufgrund der bereits gemeldeten Fälle der letzten zwei Wochen vorhergesagt werden, dass sich die Zahl der wöchentlichen Todesfälle fast verdoppeln wird. Prognosen über mehr als zwei Wochen hinaus gestalten sich als schwierig, da die Zahl der Todesfälle von den noch nicht gemeldeten Fällen abhängt. Um diese Unsicherheit einzugrenzen, setzen wir zwei Zukunftsszenarien an, wobei wir in (A) einen linearen Anstieg und in (B) einen exponentiellen Anstieg der Fälle annehmen (Methoden in den ``supplementary material''). Die sich daraus ergebenden Vorhersagen bewegen sich zwischen 500 und 800 wöchentlichen Todesfällen Anfang November. Die Fallzahlen könnten jedoch noch schneller ansteigen, falls sich die zweite Welle voll entfaltet. 

Wenn die Kapazitätsgrenze für Testung und Kontaktnachverfolgung (TTI) überschritten ist, brechen die Eindämmungsmaßnahmen dieses Kontrollsystems zusammen und der Anstieg der Fallzahlen beschleunigt sich~\cite{contreras_challenges_2020}.
Die aktuelle Entwicklung in Deutschland bringt somit die in der Vergangenheit erfolgreiche Bewältigung der Pandemie ernsthaft in Gefahr. Um die Kontrolle wieder zu erlangen und um ein Überschreiten des Kipppunktes bei Überlastung der TTI-Kapazität zu vermeiden, müssen die Fallzahlen unverzüglich gesenkt werden. Andernfalls werden die Eindämmung der Ausbreitung und der Schutz der Risikogruppen zwangsläufig sehr viel restriktivere Maßnahmen erfordern -- spätestens wenn die Krankenhauskapazität erreicht ist.

\begin{table}[H]
    \centering
    \vspace{1cm}
    \begin{tabular}{l|l|l|l|l||l|l}
    \multicolumn{2}{r|}{Anzahl}&\multicolumn{3}{c||}{Anzahl} &\multicolumn{2}{c}{}\\
    \multicolumn{2}{r|}{Wochen}&Fälle&\multicolumn{2}{c||}{Tode}& CFR &$R$\\
Zeitraum&&&prog.&beob.&(in \%)&(RKI) \\ \hline
Juli&4&14163 & 122 & 146 &1,0&1,11\\
August&4&32273 & 156 & 127&0,39&1,06\\
Anf. Sep&3& 30493 & 214 & 184 &0,6&1,10\\
Ende Sep&2& 28737 & 281 & 286 &0,99&1,16\\ 
Anf. Okt&2& 66796 & 741 &- & -& 1,31\\
    \end{tabular}
    \caption{\textbf{Vergleich der prognostizierten und beobachteten Todesfälle. } Diese Tabelle fasst die in (Abb.~\ref{fig:DE_DE} vorgestellten Zahlen zusammen. Für jeden Zeitraum im Sommer 2020 werden die Anzahl der gemeldeten COVID-19 Neuinfektionen (``Fälle") zusammen mit der Anzahl der Todesfälle aufgeführt, die beobachtet (beob.) und auf Basis von \cite{Levin2020} prognostiziert (prog.) werden. Um dem zeitlich verzögerten Zusammenhang zwischen Fall- und Todesmeldung Rechnung zu tragen, werden alle Todesfälle zwei Wochen früher terminiert. Die prognostizierten Todesfälle weichen von den beobachteten Todesfällen um weniger als 20\% ab. Die Fallsterblichkeit (CFR) wird als Verhältnis der beobachteten Todesfälle zu den gemeldeten Fällen berechnet. Zum Vergleich: Wenn die Fälle gleichmäßig über die Altersgruppen verteilt wären und keine Fälle unentdeckt blieben, würde die CFR der theoretischen Infektionssterblichkeit (IFR) entsprechen, die für Deutschland bei 1,61 liegt mit einem 95\%-Konfidenzintervall von [1,29;2,02]~\cite{Levin2020}. Die letzte Spalte zeigt die Reproduktionszahl R, wie vom Robert Koch-Institut (RKI) gemeldet.
    \label{tab:IFR_CFR_DE}
    }
\end{table}

\end{document}